\documentclass[aps,twocolumn,prb]{revtex4}
\usepackage{graphicx}
\usepackage{amsmath}
\usepackage{amsfonts}
\begin{document}
\title{Lattice-gas modeling of CO adlayers on Pd(100)}
\author{Da-Jiang Liu}
\affiliation{Ames Laboratory (USDOE), Iowa State University, Ames, Iowa 50011}
\date{\today}

\begin{abstract}

Using a lattice-gas model with pairwise interactions, we study the
ordered structures, coverage dependence of the heat of adsorption, and
other experimentally observable behavior of adsorbed CO overlayers on
Pd(100) single crystal surfaces.  Transfer matrix and Monte Carlo
methods give accurate information regarding the lattice-gas model that
often contradicts simple mean-field-like analysis.  We demonstrate the
usefulness of the model by reproducing experimental results over a
large range of pressures and temperatures.
\end{abstract}

\maketitle

\section{Introduction}

CO adsorption on metal surfaces has been studied extensively as a
benchmark system for chemisorption.  In particular, great deal of
information has been accumulated during the last thirty years about CO
adsorption on Pd(100) surfaces using several different experimental
techniques, and under a range of pressures and
temperatures.\cite{tracy69,bradshaw78,behm80,szanyi93} There are also
detailed theoretical studies of CO adsorption on Pd(100) using
first-principles approaches.\cite{eichler98,wu00} However,
surprisingly, no extensive statistical mechanics studies have been
performed to precisely determine adlayer ordering and phase
transitions for this system.  This is a significant omission since
such analysis provides strong constraints on the type and magnitude of
adspecies interactions. It also provides a reliable determination of
thermodynamic quantities, which is not possible with simplified
analyses. Such information is invaluable in interpreting other
experiments, e.g., related to CO adsorption energies.

A long-standing motivation for such detailed studies of simple
chemisorption systems is to provide insight into catalytic surface
reactions. It is also well-recognized that ordering and islanding of
reactants will limit the utility or validity of mean-field type rate
equation treatments of the reaction kinetics.\cite{wintterlin97}
Hence, accurate and robust atomistic modeling of adlayer structure for
individual reactants is a crucial first step in building realistic
atomistic model of related surface reactions, e.g., for CO oxidation
on Pd(100).\cite{liu03c}

In this paper, we develop and analyze a lattice-gas (LG) model for CO
adlayers on Pd(100).  Our focus is on equilibrium aspects of this
system, since molecularly adsorbed CO can diffuse quite rapidly 
on this surface under normal situations facilitating adlayer
equilibration.  However, some nonequilibrium
issues are also addressed.  The main techniques used to
analyze behavior of the lattice-gas model are the
transfer matrix method and Monte Carlo simulation.  We aim to
reproduce as many experimental observations as possible using a
relatively simple model.  We present results regarding
surface ordering below 0.5 monolayers (ML) in Sec.~\ref{sec:ordering}, 
the structure of dense CO adlayers in Sec.~\ref{sec:comm-incomm-trans},
the heat of adsorption in Sec.~\ref{sec:heat-adsorption},
and the adsorption isobars in Sec.~\ref{sec:isobars}. 

\section{Lattice-gas model for CO/Pd(100) and its analysis}

Various experiments\cite{tracy69,ortega82,szanyi93} show that adsorbed
CO resides only at bridge sites on Pd(100). Thus, our modeling of
equilibrated adlayer configurations allows population of bridges sites
only. We note, however, that it was suggested\cite{eichler98} that
during adsorption, CO is first steered towards the less favorable top
sites. Thus, in more general modeling of nonequilibrium configurations
under reaction conditions, it is appropriate to allow population of
other sites.\cite{liu03c} Below, $a = 2.75$ \AA denotes the
unit cell size of Pd(100) surface.

\begin{figure}[!tbp]
\includegraphics[width=2.5in]{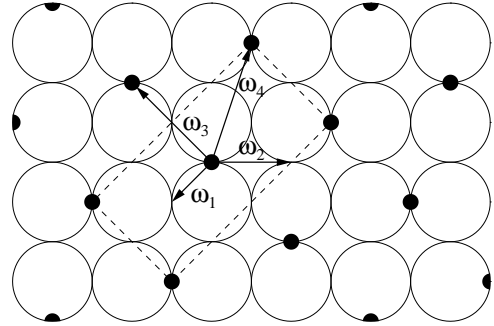}
\caption{Schematic of the $c(2\sqrt{2} \times \sqrt{2})R45^\circ$
  ordered structure and of the pairwise interactions used in this
  paper. The large open circles represent Pd atoms and the small solid
  circle represents CO molecules adsorbed on the bridge site of the
  Pd(100) surface.}
\label{fig:schem_coint}
\end{figure}

Our LG modeling also assumes only pairwise-interactions between CO
adsorbates.  Figure~\ref{fig:schem_coint} illustrates the specific
interactions used: we incorporate nearest-neighbor (NN) interactions
$\omega_1$ for CO pairs separated by distance $a/\sqrt{2}$, second NN
(2NN) interactions $\omega_2$ for separation $a$, third NN (3NN)
interactions $\omega_3$ for separation $\sqrt{2} a$, and sometimes
fourth NN (4NN) interactions $\omega_4$ for separation $\sqrt{5/2} a$.
Also illustrated in the figure is the experimentally observed
$c(2\sqrt{2} \times \sqrt{2}) R 45^\circ$ ordered structure.

In applying our model to analyze the heat of adsorption and the
adsorption isobars for CO/Pd(100), we need also information of the
initial heat of adsorption at low coverage, which corresponds to the
absorption energy of an isolated CO molecule at a bridge site.  This
quantity was measured at 1.55 eV in an early experiment by Tracy and
Palmberg,\cite{tracy69} and at 1.67 eV in more recent
experiments.\cite{behm80,szanyi93} Using density functional theory
(DFT) incorporating the generalized gradient approximation (GGA),
Eichler and Hafner\cite{eichler98} studied the potential energy
surface for CO adsorption on Pd(100).  They reported the adsorption
energy for CO on a bridge site to 1.92 eV.  This value is larger than
the experimental estimate, but the trend that bridge sites are favored
over top and hollow sites is consistent with experimental findings.

To analyze the above two-dimensional LG model, in this paper we use
two standard yet powerful statistical mechanical techniques: the
transfer matrix (TM) and the Monte Carlo (MC) methods. These two
methods are often complementary.  Using TM, one can always obtain the
equilibrium free energy and other thermodynamic properties of the
system.  However, it is difficult to include long-range interactions
using TM.  For example, performing analysis on a system or strip of
size $M \times \infty$, in order to include 3NN interaction
$\omega_3$, it is necessary to consider all configurations of two
columns spanning the strip. Thus, one must consider a total of
$2^{2M}$ configurations, if one does not reduce this number by
exclusion and symmetry properties. On the other hand, it is relatively
straight-forward to include long-range interactions using the MC
method.  However, standard MC can become inefficient, especially when
there are strong repulsive adspecies interactions, or for low
temperatures.

\section{$c(2\sqrt{2} \times \sqrt{2}) R45^\circ$ ordering below 0.5 ML}
\label{sec:ordering}

From the observation of $c(2\sqrt{2} \times \sqrt{2}) R45^\circ$
ordered structure, it has been deduced that lateral interactions
between CO(ads) consist of strong NN and 2NN repulsions, and a weak
3NN repulsion.  It has also been pointed out by Behm \textit{et
al.}\cite{behm80} that the 3NN repulsion $\omega_3$ should not be too
strong, otherwise it would instead produce a $(\sqrt{5/2} \times
\sqrt{5/2})R18.4^\circ$ structure, which is not observed
experimentally.

In order to quantify the effect of the 3NN interaction on the ordering
of CO adsorbates, we conduct a study of the phase transitions of the
lattice-gas model with very strong NN and 2NN repulsions (i.e.,
exclusion), and a finite 3NN repulsion ($\omega_3 >0$).  Different
aspects of this model has been studied previously, but the complete
picture is not available.

In the case of no 3NN interactions ($\omega_3=0$), it was found that
there exists a very ``weak'' transition from a disordered phase to a
semi-ordered phase upon increasing the CO coverage above about
$\theta_\mathrm{CO}=0.477$.\cite{ree67} At the maximum
$\theta_\mathrm{CO}=0.5$, the semi-ordered phase consists of
alternating half-filled diagonal rows of bridge sites which can slide
with respect to each other without energy penalty.  In the limit of
very strong 3NN repulsion ($\omega_3=\infty$), it was
found\cite{bellemans66} that there is a first-order liquid-solid-like
transition with increasing CO coverage. The ordered phase has a
$(\sqrt{5/2} \times \sqrt{5/2}) R18.4^\circ$ structure.

(Traditionally, results of the above studies are described with
respect to the square lattice of bridge sites. This lattice is
rotated by 45$^\circ$ from the square lattice of Pd(100) substrate
atoms, and has twice the number of sites as Pd atoms. Thus,
the coverage on this lattice satisfies $\rho=0.5\theta_\mathrm{CO}$. 
For $\omega_3=0$, the semi-ordered phase is commonly denoted as (2$\times$1)
phase, although it has long-range order in one dimension only.
It has a maximal coverage of $\rho=0.25$. For $\omega_3=\infty$, the ordered
phase is described as $(\sqrt{5} \times \sqrt{5}) R26.6^\circ$ with
respect to the square lattice of bridge sites.)

Our TM and MC study shows that as $\omega_3$ increases from zero, the
``weak'' transition to the semi-ordered phase becomes stronger, and
for larger 3NN repulsion, the transition converts to a first-order
transition to the $c(2\sqrt{2} \times \sqrt{2}) R45^\circ$ ordered
structure. This implies the existence of a tricritical point.  From MC
simulations, we estimate that this tricritical point is located at
$(\beta \mu^t, \beta \omega_3^t)=(5.2, 0.3)$, where $\beta=1/(k_BT)$
and $\mu$ denotes the chemical potential for the adsorbed CO (which is
described in more detail in the following sections). The corresponding
coverage is $\theta_\mathrm{CO}=0.46$ (or $\rho=0.23$).

\begin{figure}[!tbp]
\includegraphics[width=3.4in]{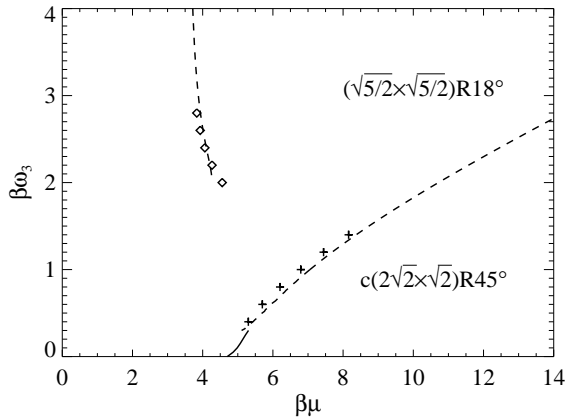}
\caption{$\mu$-$\theta$ phase diagram of the lattice-gas model with NN
  and 2NN exclusion and 3NN repulsion.  The solid line denotes a
  continuous transition obtained from transfer matrix (TM)
  finite-size-scaling\cite{nightingale76} (10-12 scaling).  Also shown
  are two first-order transitions obtained from locating local maximum
  in $d \theta / d \mu$ using two methods: the dashed lines are from
  TM (with strip of size 10) calculations, and the symbols are from
  Monte Carlo simulations with the histogram method.\cite{ferrenberg88}}
\label{fig:tm2d_phase_xmu_e3}
\end{figure}

As $\beta \omega_3$ further increases above approximately 1.9, there
is another first-order transition, but this time, the transition is
from the disordered phase to a $(\sqrt{5/2} \times \sqrt{5/2}) R
18.4^\circ$ phase.  Figure~\ref{fig:tm2d_phase_xmu_e3} shows our
preliminary results directed towards mapping out the $\mu$-$\beta$
phase diagram of the model with NN and 2NN exclusion, and 3NN
repulsion.  It shows two first-order liquid-solid-like transitions
(dashed lines) from disordered phase to either a $c(2\sqrt{2} \times
\sqrt{2}) R45^\circ$ ordered phase, or a $(\sqrt{5/2} \times
\sqrt{5/2}) R 18.4^\circ$ ordered phase.  Also there is a
solid-solid-like transition from the $(\sqrt{5/2} \times \sqrt{5/2}) R
18.4^\circ$ phase to the $c(2\sqrt{2} \times \sqrt{2}) R45^\circ$
phase.  There is a tricritical point connecting the continuous and
first-order transitions to the $c(2\sqrt{2} \times \sqrt{2})
R45^\circ$ phase.  However, how the termination of the $(\sqrt{5/2}
\times \sqrt{5/2}) R 18.4^\circ$ transition line is not clearly
determined from the existing analysis.  It is likely that the
transition line bends towards and merges with the $c(2\sqrt{2} \times
\sqrt{2}) R45^\circ$ transition line sharply at around $\beta \omega_3
= 1.9$.

Under the framework of the lattice gas model with only repulsive
interactions between neighboring pairs up to 3NN, one can conclude
that $\omega_3 < 0.5$ eV so that at room temperature, no $(\sqrt{5/2}
\times \sqrt{5/2}) R 18.4^\circ$ phase shall be observed.  A caveat is
that introducing further neighboring interactions can change the phase
diagram significantly and the above constraint on the magnitude of
$\omega_3$ is no longer valid if longer-ranged interactions should be
considered.

Based on the above analysis (and also our investigations in subsequent
sections), we assign a value of $\omega_3 = 0.03$ eV for the strength
of the 3NN repulsive interactions.  In the following analyses which
include consideration of behavior for CO coverage above 0.5 ML, it is
necessary to relax the constraint of 2NN exclusions. The value of
$\omega_2$ of around 0.17 eV is determined from our analysis of the
heat of adsorption in Sec.~\ref{sec:heat-adsorption}, and adsorption
isobars in Sec.~\ref{sec:isobars}. However, before presenting these
analyses, in Sec.~\ref{sec:comm-incomm-trans}, we provide a more
complete picture of adlayer ordering by describing behavior at
coverage above 0.5 ML (using the parameter choice $\omega_1 =
\infty$, $\omega_2 = 0.17$ eV, and $\omega_3 = 0.03$ eV.)

\section{Structure of dense CO adlayers}
\label{sec:comm-incomm-trans}

Because of the difficulties of experimental techniques (e.g., work
function and infrared analysis) in dealing with high CO coverage
($\theta_\mathrm{CO} > 0.5$), structures of dense CO adlayers on
Pd(100) are a matter of some debate.\cite{uvdal88,berndt92,schuster96}
As $\theta_\mathrm{CO}$ increases above 0.5 ML, it is concluded from
diffraction studies that the adlayers structure undergoes a
commensurate-incommensurate transition (CIT).  The study by Schuster
\textit{et al.}\cite{schuster96} suggests it is in the
Pokrovsky-Talapov universality class, consistent with expectation from
symmetry arguments.

Lattice-gas models are not ideal for study of CIT's, since they put
too many constraints on the structure of domain walls.  Nonetheless,
we perform a Monte Carlo study to investigate CO adlayer structure
above 1/2 ML.  We simulate the system at a fixed pressure while
lowering the temperature.  We use Glauber dynamics (corresponding to
adsorption/desorption in the LG model) with the Metropolis algorithm.
Occasionally, we also mix in Kawasaki dynamics (corresponding to
diffusion in the LG model) with the Glauber dynamics.  Here, our focus
is in the equilibrium structure of CO adlayer, thus we do not need
to mimic the physical kinetics accurately. The primary challenge is
that due to CO adspecies repulsions, adlayers can become nearly
``frozen'' using normal dynamics at high CO coverage and low
substrate temperature.  In fact, the LG model with NN and 2NN
exclusion has been used to study the glass transition.\cite{rao92}

\begin{figure}[!tbp]
\includegraphics[width=3.0in]{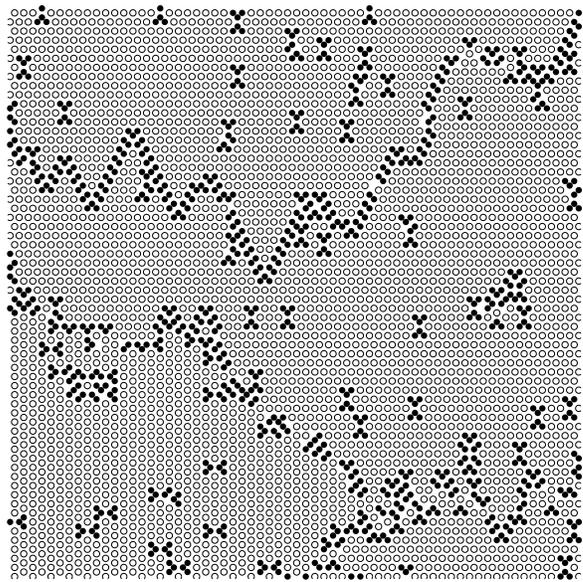}
\caption{Snapshot of Monte Carlo simulations using the Glauber
  dynamics and the Metropolis algorithm for the LG model at fixed
  pressure while lowering the temperature.  The annealing rate is
  $10^4$ MCS/K (each site in the system is sampled once on average for
  each MCS).  Other parameters are $p_\mathrm{CO}=10^{-7}$ Torr,
  $\omega_1=\infty$, $\omega_2=0.17$ eV, $\omega_3=0.03$ eV.  The
  snapshot is taken at $T=400$ K.  For illustration, we denote CO with
  exactly two 2NN and four 3NN by a circle, and all other CO
  (``defects'') by a black dot.  Also note that the Pd(100) substrate
  (not shown) is rotated 45$^\circ$.  Shown in the figure is a $L=128$
  subsystem in a simulation with $L=256$ using periodic boundary
  conditions. $\theta_\mathrm{CO}=0.511$ ML.}
\label{fig:mc2b256_isobar_site}
\end{figure}

Although our simulations are not faithful to the physical adlayer
dynamics, they provide at least some qualitative insights into adlayer
structure.  For a fixed system size with periodic boundary conditions,
if we lower the temperature very slowly, eventually a single domain
occupies the whole system.  Upon further lowering the temperature,
defects are formed.  However, we are unable to observe any
well-defined domain wall structure as suggested by Berndt and
Bradshaw.\cite{berndt92} On the other hand, if we lower the
temperature more quickly, then different domains still occupy the
system near the transition point.  Further decreasing the temperature
is accompanied by the enlargement of those original domains, and
emergence of defects inside different domains.  We show in
Fig.~\ref{fig:mc2b256_isobar_site} a snapshot of such a configuration
generated by Monte Carlo simulations.

\section{heat of adsorption}
\label{sec:heat-adsorption}

Assuming equilibrium between CO in the gas phase and the chemisorbed
phase, the Clausius-Clapeyron equation relates the gas phase pressure,
$P$, to an isosteric heat of adsorption, $E_\mathrm{st}$, via
\begin{equation}
[d \ln P/d(1/T)]_\theta = -E_\mathrm{st}/k_B.
\label{eq:cc}
\end{equation}
Various experimental techniques\cite{tracy69,behm80,szanyi94,yeo97}
give similar results for the value of $E_\mathrm{st}$ as
$\theta_\mathrm{CO} \to 0$, while conflicting results have been
obtained for the coverage dependence of $E_\mathrm{st}$.  Most
studies\cite{tracy69,szanyi94,yeo97} show a decrease in
$E_\mathrm{st}$ with increasing $\theta_\mathrm{CO}$, while the study
by Behm \textit{et al.}\cite{behm80} shows a roughly constant
$E_\mathrm{st}$ for $\theta_\mathrm{CO} < 0.5$.  Presence of carbon is
suggested\cite{behm80} as the reason for this discrepancy, a claim
disputed by others.\cite{yeo97}

The presence of lateral interactions between CO adspecies is often
invoked\cite{tracy69,yeo97} to explain the coverage dependence of
$E_\mathrm{st}$.  Yet to our knowledge, no systematic study of the
isosteric heat of adsorption in an interacting lattice-gas model for
this system has been performed previously. There are some general
studies of effects of lateral interactions on the heat of adsorption
using mean-field approximations (see, e.g.,
Ref.~\onlinecite{al-muhtaseb99} and references therein) which are not
reliable for models with relatively short-range interactions.
 
In lattice-gas modeling, it is appropriate to use the grand canonical
ensemble.  We assume simply that the gas phase pressure $P$ is related
to the chemical potential through $\mu = k_B T \ln (P/P_0)$.  More
accurate forms of the relationship between the pressure and the
chemical potential will introduce corrections to the heat of
adsorption on the order of $k_B T$, which is negligible for present
purposes. Unlike the isobar experiments (treated later in
Sec.~\ref{sec:isobars}), for our analysis here, we do not need
information regarding prefactors for desorption, or the sticking
coefficient for adsorption.

\begin{figure}[!tbp]
\includegraphics[width=3.4in]{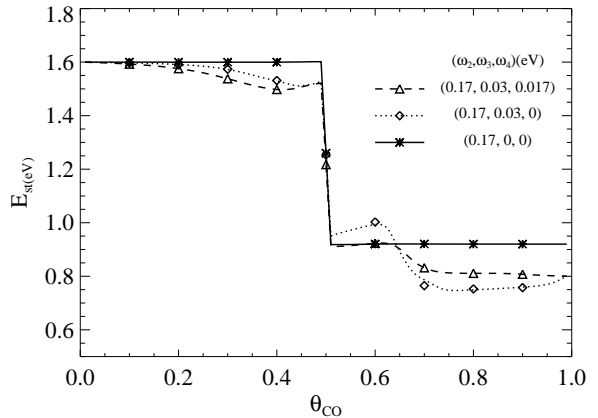}
\caption{Transfer matrix calculation of the isosteric heat of
adsorption versus CO coverage on Pd(100).  The lines are obtained
using strips of size 8, and the symbols are obtained using strips of
size 10.  Nearest-neighbor interaction $\omega_1$ is assumed to be
infinite and other neighboring interactions are shown in the
figure. $T=300$ K.}
\label{fig:tm2g8_isosteric}
\end{figure}

Using the transfer matrix method, we examine the coverage dependence
of $E_\mathrm{st}$ for the lattice-gas model with various choices of
interactions.  Any coverage dependence would reflect the influence
of adspecies interactions which can cause $E_\mathrm{st}$ to deviate
from its limiting value of $\epsilon_b$ for low coverage. Some of the results
are shown in Fig.~\ref{fig:tm2g8_isosteric}.

\begin{figure}[!tbp]
\includegraphics[width=2in]{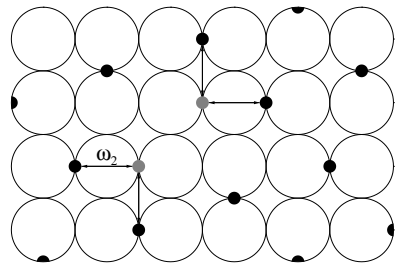}
\caption{Schematic showing the accommodation of an extra
CO molecule in an otherwise perfect $c(2\sqrt{2}
  \times \sqrt{2}) R 45^\circ$ structure.}
\label{fig:heat_schem2}
\end{figure}

For the lattice gas model with only 2NN repulsive interaction only,
the heat of adsorption is effectively a step function, with the
form\begin{equation} E_\mathrm{st} \approx
\begin{cases}
\epsilon_b & \text{if $\theta < 0.5$} \\
\epsilon_b - 4 \omega_2 & \text{if $0.5 < \theta < 1$}. 
\end{cases}
\end{equation}
The result can be explained as follows: for $\theta < 0.5$, adsorbed
CO molecules can rearrange themselves to avoid any 2NN pairs.  Around
$\theta=0.5$, they form a near-perfect $c (2 \sqrt{2} \times \sqrt{2})
R 45^\circ$ adlayer.  To adsorb more CO molecules beyond this
near-perfect overlayer, it is necessary to create four 2nd NN pairs.
See Fig.~\ref{fig:heat_schem2} for an illustration.

With longer-ranged repulsive interactions, $E_\mathrm{st}$ decreases
as $\theta_\mathrm{CO}$ increases even for $\theta_\mathrm{CO} <
0.5$.  However, the coverage dependence is quite nonlinear.  For
example, with 3NN interactions only (the dotted line), $E_\mathrm{st}$
only decreases slightly for $\theta_\mathrm{CO} < 0.25$.  This is
again due to the fact that below this coverage, CO(ads) can easily
arrange themselves in a way to avoid any 3NN pairs.  

With the present set of parameters for lateral interactions, the
lattice-gas model produces a coverage dependence of the heat of
adsorption somewhat between the experimental results of Behm
\textit{et al.} and other groups.  Most of the decrease in the heat of
adsorption occurs when $\theta_\mathrm{CO} > 0.5$, while only a
slight decrease occurs when $\theta_\mathrm{CO} < 0.5$.

The near parabolic decrease in $E_\mathrm{st}$ starting from
$\theta_\mathrm{CO}=0$, as well as the transient increase after the
ordering transition, shown by Fig.~\ref{fig:tm2g8_isosteric} are quite
reminiscent to the experimental results of Guo and Yates\cite{guo89}
for CO adsorption on Pd(111). However, they reported a plateau at an
ordering transition, while we see first a sharp drop and then a
plateau at the ordering transition point.

\section{Adsorption Isobars}
\label{sec:isobars}

For a CO adlayer in equilibrium with gas phase CO, it is clear that as
the surface temperature increases (at fixed pressure), the CO coverage
will decrease. For higher (fixed) pressures, this decrease will be
delayed until a higher temperature range. To quantify this behavior
using our LG model, one needs a more quantitative determination of the
relationship between pressure and chemical potential than that
presented in the previous section.  To this end additional assumptions
are needed.  We assume that the impingement rate is given by
$P/\sqrt{2 \pi m k_B T}$ and the attempt frequency for desorption is
$\nu_0$, then we assume that
\begin{equation}
\mu = E_\mathrm{st} + k_B T \ln \frac{P}{\nu_0 \sqrt{2 \pi m k_B T}}.
\end{equation}
The initial sticking coefficient when $\theta_\mathrm{CO}=0$ is
taken to be unity.  Thus by assuming equilibrium of CO between gas
phase and the chemisorption phase, one can calculate the adsorption
isobar of the lattice gas model using either the transfer matrix or
the Monte Carlo method.

The results are shown in Fig.~\ref{fig:tm2g_isobar}.  For low
pressures, as $T$ decreases, $\theta_\mathrm{CO}$ first increases,
then reach a plateau at $\theta_\mathrm{CO}=0.5$.  For
$p_\mathrm{CO}=10^{-7}$ Torr, the plateau occurs near $T=420$ K.  This
result is in very good agreement with
experiments.\cite{behm80,szanyi94} Upon a further decrease in
temperature, $\theta_\mathrm{CO}$ again increases above 0.5 ML.
Experimentally, this corresponds to the adlayer moving into the regime
of the commensurate-incommensurate transition discussed in
Sec.~\ref{sec:comm-incomm-trans}.  Here, we simply note that the
temperature where this occurs depends sensitively on the value of the
2NN interactions $\omega_2$.  By choosing $\omega_2=0.17$ eV, the
transition occurs between 340 K to 400 K for $p_\mathrm{CO}$ between
$10^{-9}$ to $10^{-7}$ Torr, in agreement with
experiments.\cite{szanyi94,schuster96} It is also significant that at
high pressures, the LG model predicts disappearance of the plateau
near $\theta_\mathrm{CO}=0.5$, which is also observed
experimentally.\cite{szanyi94}

\begin{figure}[!tbp]
\caption{Adsorption isobars calculated from the lattice-gas model.
  Two sets of parameters are used.  $p_\mathrm{CO}$ ranges from
  $10^{-9}$ to $0.1$ Torr.}
\label{fig:tm2g_isobar}
\end{figure}

Note that we ``naively'' choose $\nu_0 = 10^{13} \, \mathrm{s}^{-1}$
for the desorption prefactor, while Behm \textit{et al.} obtain a
value on the order of $10^{16} \, \mathrm{s}^{-1}$ from their
adsorption isobars.  Consequently there is some discrepancy between
the lattice-gas model prediction and experiments at low coverage.
Specifically for $\theta_\mathrm{CO} < 0.5$ ML, significant desorption
occurs at a higher temperature than experiments.  Adopting adsorption
energy $\epsilon_b$ from DFT calculations and assuming the same
prefactor makes the discrepancy even larger.  Also the discrepancy
will not likely be resolved by modification of interactions, since at
low coverage lateral interactions are quite insignificant.

\bigskip

\section{Summary}
\label{sec:summary}

We have performed a combined transfer matrix and Monte Carlo study of
a lattice-gas model for CO adlayers on Pd(100). Model predictions are
compared against a variety of experimental observations. Of particular
significance is our determination of repulsive adspecies interactions:
$\omega_1=\infty$, $\omega_2=0.17$ eV, and $\omega_3=0.03$ eV.  Our
estimate of the 2NN interaction of 0.17 eV agrees well with Wu and
Metiu.\cite{wu00} Our assignment of a weak 3NN interaction is
consistent with early qualitative arguments by Behm \textit{et
al.}\cite{behm80} Our use of a binding energy of $\epsilon_b=1.6$ eV in
analysis of the heat of adsorption and the adsorption isobars is
consistent with experimental estimates, but this value is somewhat
smaller than that from DFT which is close to 2 eV.

\acknowledgments

The author thank Profs. T. L. Einstein and P. A. Thiel for helpful
discussions, and Prof. J. W. Evans for extensive discussion and
suggestions.  This work is supported by the Division of Chemical
Sciences, U.S.  Department of Energy (USDOE). It was performed at Ames
Laboratory which is operated for the USDOE by Iowa State University
under Contract No. W-7405-Eng-82.

\end{document}